\documentclass[prb,preprint,draft,amsmath,showpacs]{revtex4}
\usepackage{graphics}

\begin{document}

\bibliographystyle{prsty}
\input epsf

\title {Optical investigation of the charge-density-wave phase 
transitions
in $NbSe_{3}$}

\author {A. Perucchi$^{1}$, L. Degiorgi$^{2,1}$, and R.E. 
Thorne$^{3}$}
\affiliation{$^{1}$Laboratorium f\"ur Festk\"orperphysik, ETH 
Z\"urich,
CH-8093 Z\"urich, Switzerland}\
\affiliation{$^{2}$Paul Scherrer Institute, CH-5232 Villigen}\
\affiliation{$^{3}$Department of Physics, Cornell University, Ithaca
NY 14853, U.S.A.}

\date{\today}

\begin{abstract}
We have measured the optical reflectivity $R(\omega)$ of the quasi
one-dimensional conductor $NbSe_{3}$ from the far infrared up to the 
ultraviolet between 10 and 300 $K$ using light polarized along and normal to 
the chain axis. We find a depletion of the optical conductivity with 
decreasing temperature for both polarizations in the mid to 
far-infrared region.  This leads to a redistribution of spectral 
weight from low to high energies due to partial gapping of the Fermi 
surface below the charge-density-wave transitions at 145 K and 59 K.  
We deduce the bulk magnitudes of the CDW gaps and discuss the 
scattering of ungapped free charge carriers and the role of 
fluctuations effects.

\end{abstract}
\maketitle

When metals are cooled, they often undergo a phase transition to a
state characterized by a new type of order. Of particular interest are
the quasi-one-dimensional linear-chain metals because they exhibit 
important
deviations from Fermi liquid behaviour and unusual phenomena
associated with charge- and spin-density-wave broken symmetry ground 
states \cite{gruner,vescoli}.
A charge-density-wave
(CDW) is a condensate comprised of a coupled modulation of the 
conduction electron density and lattice atom positions.  It forms via 
a Peierls
transition, in which an instability of the metallic Fermi surface 
(FS) due to nesting at $q=2k_{F}$ ($k_{F}$ being the Fermi 
wave-vector) couples to the Kohn anomaly in
the phonon spectrum \cite{pouget89,pretransflucts,fleming}.  As in conventional 
superconductivity the electron-phonon coupling is the 
dominant interaction \cite{gruner}, and the transition leads to 
opening of a charge gap. 

The transition metal trichalcogenides $MX_{3}$ with $M=Nb, Ta, Ti$ 
and $X=S, Se, Te$ are among the most interesting materials displaying 
low dimensional
electronic properties. Their crystallographic structure is made up of
infinite chains of trigonal prisms \cite{gruner}. Depending on the
coupling between chains, they can exhibit pseudo-gap 
one-dimensionality
to anisotropic three dimensionality. The most remarkable properties 
have been observed in $NbSe_{3}$, which shows the phenomena of density 
wave transport cleanly than any other known system \cite{gruner}. The resistivity
remains metallic down to low temperatures, but two CDW phase 
transitions occur at
$T_{1}=145 ~K$ and $T_{2}=59 ~K$ where the resistivity shows a sharp
increase \cite{ong77}. Tight binding calculations
\cite{canadell} have shown that $NbSe_{3}$ has a primarily
one-dimensional character with deviations due to the short intra- and 
interlayer $Se-Se$ contacts.  This transverse coupling produces a 
warped Fermi surface so that imperfect nesting and only partial 
gapping are expected.   

The transition at $T_{1}$ is associated with a linear
nesting, the wavevector of the CDW condensate pointing along the
main chain axis, and the one at $T_{2}$ with a diagonal nesting
\cite{schaefer}. The literature values for the energy gaps, obtained 
by surface-sensitive techniques including point-contact, tunneling 
and angle
resolved photoemission (ARPES) spectroscopies 
\cite{ekino,dai,fournel,monceau,schaefer2}, vary 
significantly:
$2\Delta_{1}\sim 110-220 ~meV$ for the $T_{1}$ CDW, and
$2\Delta_{2}\sim 40-90 ~meV$ for the $T_{2}$ CDW.
In all cases the gap values yield a ratio $2\Delta_{i}/k_{B}T_{i}$ 
much larger than the mean-field value of 3.52, and the 
mean-field transition temperature $T_{MF}$ is much larger than the 
measured
$T_{1}$ and $T_{2}$. These discrepancies from mean-field values 
indicate the importance of one-dimensional fluctuations. 

In principle, optical spectroscopic methods are an ideal
bulk-sensitive tool in order to investigate the CDW phase transitions
\cite{gruner}. Previous optical spectroscopy measurements were performed at room 
temperature \cite{geserich} or over an insufficient energy range 
\cite{challaner,nakahara} which led to an erroneous evaluation of the CDW 
gaps.  We provide here the first comprehensive study of the optical
properties of $NbSe_{3}$ over a broad spectral range and as a 
function of temperature. The optical conductivity shows a 
redistribution of spectral
weight from low to high frequencies with decreasing temperature, 
which we use to determine the bulk CDW (pseudo)gaps.  We also observe 
precursor effects of the CDW phase transitions and establish that the 
resistivity anomalies are primarily the consequence of a Fermi 
surface gapping and not due to changes in the lifetime of the charge 
carriers.

High purity samples were grown as previously described \cite{thorne}. 
This study was made possible by
preparing aligned mosaic specimens consisting of several wide, flat 
$NbSe_3$ ribbons with a resulting optical surface 3 $mm$ long by 2 $mm$ wide. 
The optical reflectivity
$R(\omega)$ was measured from the far infrared up to the ultraviolet 
between $T=300 ~K$
and 10 $K$. Polarizations along and transverse to the chain axis, 
corresponding to the b and c crystallographic axes, respectively, 
were used to assess the anisotropic electrodynamic response. The 
specimens
were then coated with a 3000 $\AA$ gold layer and measured again, 
allowing correction for surface scattering from our mosaic samples 
without altering the overall shape and features of the spectra.  
Additional experimental details are described 
elsewhere\cite{vescoli,wooten}. Kramers-Kronig
transformations were used to calculate the real part
$\sigma_{1}(\omega)$ of the optical conductivity. Standard high
frequency extrapolations $R(\omega)\sim \omega^{-s}$ (with $2<s<4$)
were employed \cite{wooten} in order to extend the data set above
$10^{5}$ $cm^{-1}$ and into the electronic continuum. Because of 
scatter in the data at low frequencies (below 50 $cm^{-1}$), 
$R(\omega)$ was extrapolated using the
Hagen-Rubens (HR) law $R(\omega)=1-2\sqrt(\omega/\sigma_{dc})$ from 
data points in the 30 to 70 $cm^{-1}$ range.  This extrapolation 
yielded $\sigma_{dc}$ values in agreement with dc transport data
\cite{ong77,ong}, providing further confirmation of the 
reliability
of the gold-coating-corrected data.   The temperature dependence of 
$\sigma_{1}(\omega)$ is not
affected by the details of this low frequency extrapolation.

\begin{figure}[h]
   \begin{center}
    \leavevmode
    \epsfxsize=14cm \epsfbox {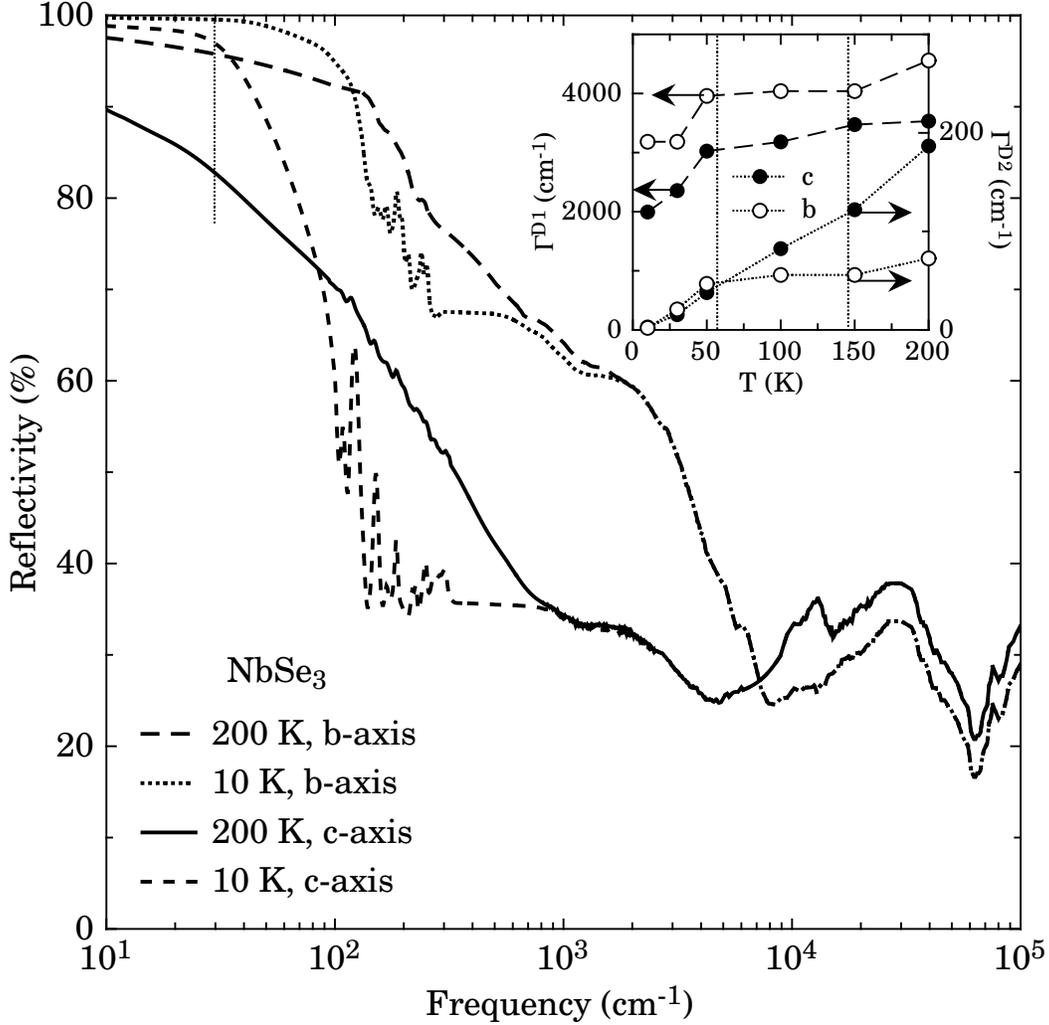}
     \caption{Optical reflectivity $R(\omega)$ in
     $NbSe_{3}$ along both polarization directions and at two selected
     temperatures (10 and 200 $K$). The thin dotted line marks the
     frequency below which the Hagen-Rubens
     extrapolation has been performed. The inset shows the 
temperature 
dependence of the scattering rates for the
     first (D1, left y-axis scale) and second (D2, right y-axis 
scale) Drude
     term in the fits to both polarization directions. Solid symbols 
refer to
     the c-axis and open ones to the b-axis. The
     CDW transition temperatures are indicated by thin dotted lines.}
\label{Refl}
\end{center}
\end{figure}

Figure 1 shows the optical reflectivity at two selected 
temperatures.  It
is metallic at all temperatures and for both polarization directions, 
but 
 there is a remarkable anisotropy between the two 
crystallographic directions. For
light polarized along the b-axis the $R(\omega)$
plasma edge has a sharp onset around 1 $eV$ ($\sim 8000 ~cm^{-1}$), 
while along the c-axis a much more gradual and broad onset begins at about 0.5 
$eV$ ($\sim 4000 
~cm^{-1}$). $R(\omega)$ along the c-axis
resembles the so-called overdamped behaviour \cite{wooten} typically seen
in low dimensional systems \cite{vescoli} and may indicate incoherent 
charge
transport along the c direction. 
Previous $R(\omega)$
data \cite{geserich} at 300 $K$ and over a smaller spectral range
also showed anisotropic behaviour. At low temperatures,
$R(\omega)$ for both polarizations is depleted in the far and
mid-infrared spectral range, but both show a sharp upturn at low 
frequencies.  This upturn leads to a crossing of the 200 $K$ and 10 
$K$ spectra
around 100 $cm^{-1}$ (well within the measured spectral range in 
Fig. 1) so that $R(\omega)$ increases with
decreasing temperatures in the $\omega\to 0$ limit. Our 10 $K$ data 
for
light polarized along the b-axis bear some similarities with earlier 
results 
 at 2 $K$ of Challaner and Richard \cite{challaner} which only 
covered the far infrared range ($\omega < 400 ~cm^{-1}$).  

\begin{figure}[h]
   \begin{center}
    \leavevmode
    \epsfxsize=12cm \epsfbox {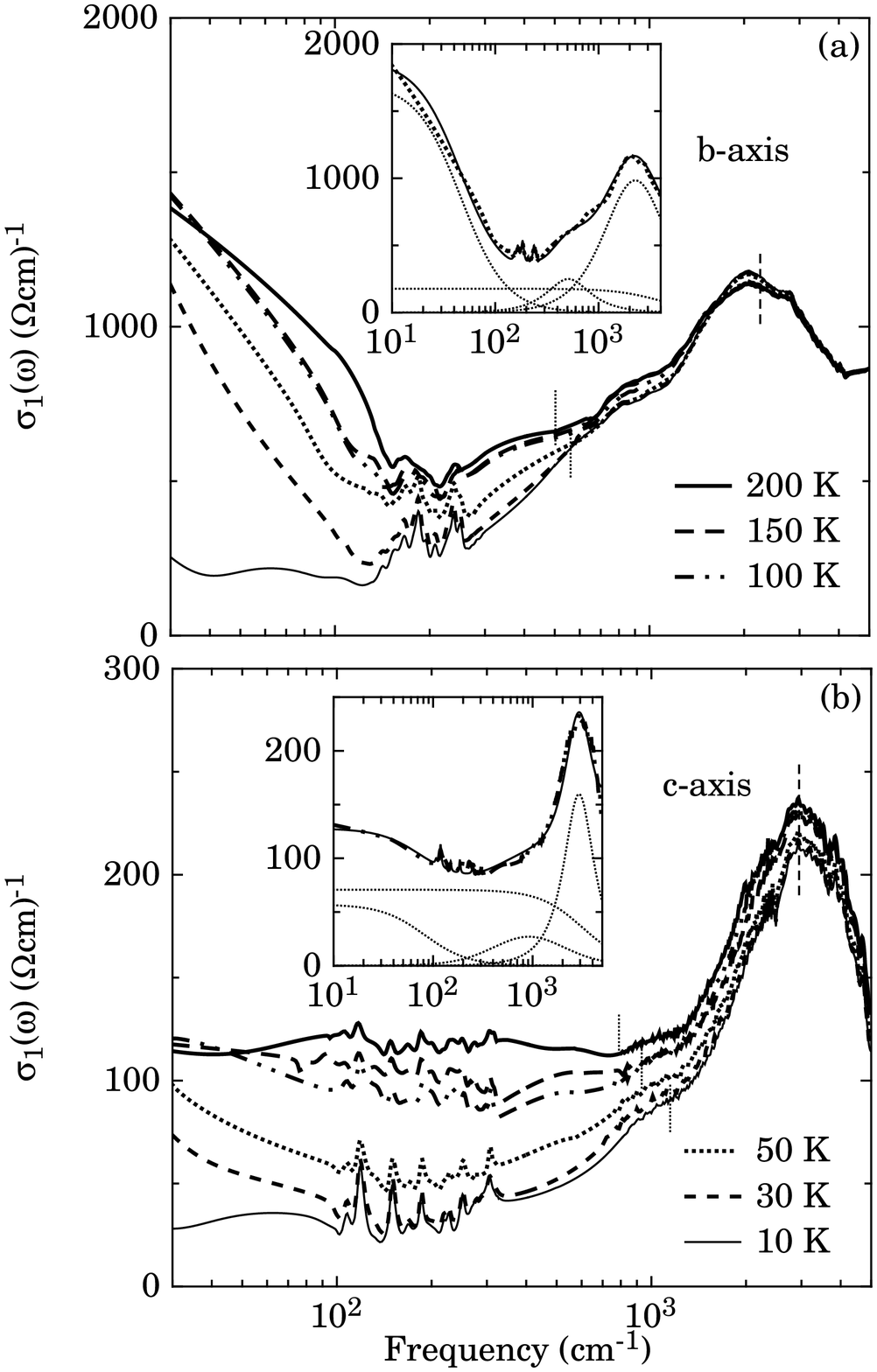}
     \caption{Real part $\sigma_{1}(\omega)$ of the optical
     conductivity above 30 $cm^{-1}$
     as a function of temperature along (a) the chain b-axis and (b) 
the
     transverse c-axis. The frequencies of the harmonic oscillators
    used to fit the two mid-infrared absorptions (see text) are
     indicated by thin dotted and dashed lines for the broad
     shoulder and the peak feature in $\sigma_{1}(\omega)$, 
respectively. The insets show
     the data at 50 (b-axis) and 100 $K$ (c-axis) with the total fit and its 
components. }
\label{sigma1/fit}
\end{center}
\end{figure}

Figure 2 shows the temperature dependence of the excitation spectrum 
below 4000 $cm^{-1}$ as revealed by the real part
$\sigma_{1}(\omega)$ of the optical conductivity. Above 4000 
$cm^{-1}$ several
absorptions (not shown here) ascribed to electronic interband
transitions \cite{geserich} are observed. In the infrared spectral 
range, $\sigma_{1}(\omega)$
for both polarizations shows a rather strong mid-infrared band at
2000 $cm^{-1}$ along the b-axis and at 3000 $cm^{-1}$ along the
c-axis. The low frequency sides of the absorptions for both the b and 
c axes have broad shoulders located below about
1000 $cm^{-1}$.
There is an obvious suppression of spectral weight in the infrared
range with decreasing temperature, while the effective (Drude) 
metallic component in the far infrared shifts to lower frequency and narrows.
The narrowing of $\sigma_{1}(\omega)$ at
low frequencies and temperatures follows from the
steep increase of $R(\omega)$ around 100 $cm^{-1}$ in Fig. 1. At
low temperatures the narrowing is so strong that the spectral weight 
of the effective (Drude) metallic component falls entirely below our data's low-frequency 
limit. This
leads to an apparent disagreement in Fig. 2 between 
$\sigma_{1}(\omega\to 0)$ and
$\sigma_{dc}$, which results from the HR extrapolation used for 
$R(\omega)$ below 50 $cm^{-1}$.    

In order to better highlight the relevant energy scales 
characterizing the
excitation spectrum and in particular to address the redistribution of 
spectral weight versus temperature in $\sigma_{1}(\omega)$ and its 
connections with the CDW transitions in $NbSe_3$, we have applied the 
phenomenological
Lorentz-Drude (LD) approach based on the classical
dispersion theory \cite{wooten}. For both polarization directions, 
the high-frequency response is fitted using Lorentz harmonic oscillators 
to describe the high-frequency electronic interband transitions.  The 
low-frequency response can be fitted with two Drude
terms and two mid-infrared harmonic oscillators at about 561 and 2270 
$cm^{-1}$ for the b-axis and
at about 1100 and 2920 $cm^{-1}$ for the c-axis (inset of Fig. 2). The two Drude terms
could represent two conduction bands \cite{ong} that are both 
affected by the 
CDW transitions. Sharp Lorentz harmonic oscillators are used to 
fit the IR-active phonon modes which appear with decreasing 
temperature \cite{challaner}. These modes contribute only a tiny 
fraction of the spectral weight below 100 $K$ in agreement with a 
previous infrared
study of $NbSe_{3}$ (Ref. \onlinecite{challaner}), and are likely 
phasons \cite{rice, degiorgi} arising from the coupling between 
the lattice and the CDW condensate.  With this minimum set of Lorentz 
and Drude terms, we obtain an outstanding fit to  
$\sigma_{1}(\omega)$  at all temperatures \cite{comment}. 

\begin{figure}[h]
   \begin{center}
    \leavevmode
    \epsfxsize=11cm \epsfbox {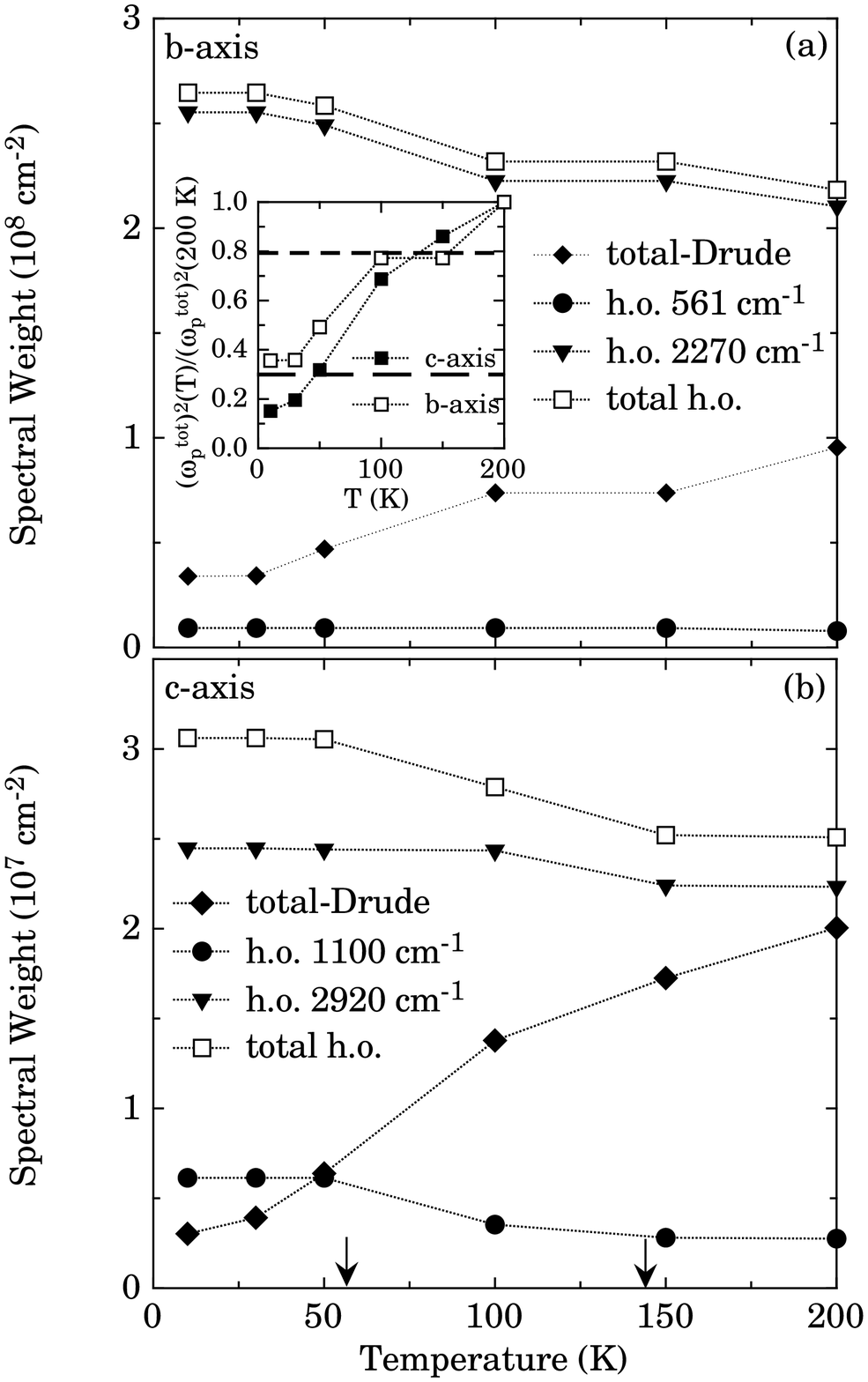}
     \caption{Temperature dependence of the total spectral weight of 
the 
     Drude terms and of the mid-infrared Lorentz harmonic oscillators, as 
well as of the spectral weights for the individual harmonic oscillators  
     (a) along the b-axis and (b) along the transverse c-axis.  The
     CDW transition temperatures are indicated by arrows. The
     inset in (a)
shows the normalized change of the total Drude weight 
$(\omega_{p}^{tot})^{2}$ with temperature, attributed to the gapping 
of the Fermi surface, for both axes. 
The dashed horizontal lines indicate the percentage of the Fermi 
surface which
survives well below each of the two CDW transitions as determined by 
magnetotransport measurements \cite{ong77}.}
\label{SW}
\end{center}
\end{figure}

Figure 3 shows how the spectral
weight is redistributed among the Drude and mid-infrared Lorentz 
terms as temperature decreases.  Both
Drude terms loose spectral weight (i.e., the
corresponding plasma frequency decreases) with decreasing 
temperature, with a more pronounced loss occurring below 100 $K$.  The 
suppressed Drude weight is transferred to high energies and in 
particular to the mid-infrared absorptions, for both polarization 
directions.  Along the b-axis the harmonic oscillator  at $\sim 2270
~cm^{-1}$ acquires most of the transferred weight, while along the
c-axis both absorptions gain weight. Part of the c-axis weight is 
transferred to even higher
energies and into the electronic continuum. Nevertheless, the total 
spectral
weight ($\int_{0}^{\infty}
\sigma_{1}(\omega)d\omega$) for
both polarization directions is fully recovered by
10$^{4}$ $cm^{-1}$ ($\sim 1 ~eV$), satisfying the optical
sum rule. In this regard $NbSe_{3}$ closely resembles the 
$2H-XSe_{2}$ dichalcogenides \cite{dordevic}, and does not show the 
"sum rule violation" of the high-temperature superconductors since 
the nature of the correlations in the two ground states are different 
\cite{dordevic,millis}.

The depletion of the Drude spectral weight in $\sigma_{1}(\omega)$ 
with decreasing
temperature indicates a progressive gapping of the Fermi surface, 
which gets partially destroyed at the Peierls transitions. The 
inset of Fig. 3a
shows the fraction change of the total Drude weight for both the b 
and c directions as a function of temperature. Ong and Monceau 
\cite{ong77}
estimated from transport data that approximately 20$\%$ of FS is 
destroyed at $T_{1}$, while 62$\%$ of the remaining 80$\%$ is 
destroyed by gaps
at $T_{2}$. The corresponding values of residual ungapped Fermi 
surface are marked by dashed lines in the inset to Fig. 3a, and are 
in excellent agreement with the present results for the b-axis. 
We attribute the 
harmonic oscillator at
2270 $cm^{-1}$ (281 $meV$) to the $T_{1}$ CDW gap \cite{gap} associated with 
linear nesting 
along the chain b-axis,
and the oscillator at 561 $cm^{-1}$ (70 $meV$) to
the $T_{2}$ CDW gap associated with the diagonal nesting 
\cite{schaefer,schaefer2}. In
c-axis data, the CDW gap due to linear nesting should
not be observable in our spectra (i.e., the transition probability is 
zero), so we attribute the harmonic oscillator at
1100 $cm^{-1}$ (136 $meV$) to the CDW gap associated with the 
diagonal nesting.  This suggests anisotropic gapping for the diagonal 
nesting\cite{monceau}. The optical gaps compare well with the broad interval of gap
estimates from tunneling, point-contact and ARPES
spectroscopy \cite{ekino,dai,fournel,monceau,schaefer2}. Our
estimated (direct) optical CDW gaps are larger than the indirect gaps 
obtained
from tunneling \cite{ekino,dai,fournel}. Deviations from the 
k-resolved ARPES findings \cite{schaefer,schaefer2} may arise in part 
because optical measurements average over the k-space.  Additional 
deviations may result because of differences between the bulk optical 
value and the near-surface value probed by these other techniques. Our 
gap value for the $T_{2}$ CDW transition and for light polarized along b is a factor of 
three to four larger than the estimate \cite{challaner} of 120-190 
$cm^{-1}$ (15-24 $meV$) obtained in the first and only other optical study at low 
temperature of $NbSe_3$.  At the time of these 
earlier measurements the large ratio 2$\Delta$/k$_B$T characteristic 
of these materials was not recognized, and so data collection did not 
extend much beyond energies corresponding to the measured $T_2$ and 
the mean-field ratio. Our c-axis $\sigma_{1}(\omega)$ data and 
 LD analysis indicate
the presence of another absorption at about 2920 $cm^{-1}$ (362 
$meV$).
This energy is too large to be ascribed to a CDW gap
\cite{ekino,dai,fournel,monceau,schaefer2,geserich, challaner}, and 
likely arises from the band structure.  It
may be a hybridization-like gap induced by Brillouin
zone backfolding arising from the Fermi surface 
nesting\cite{schaefer}.

The b and c axis absorptions at
561 and 1100 $cm^{-1}$, respectively (thin dotted lines in Fig. 2), 
slightly
soften with increasing temperature. However, the mid-infrared 
absorptions persist to temperatures well
above $T_{1}$ and $T_{2}$. This provides evidence for the importance 
of 1D fluctuations \cite{schwartz} in the regime between the measured transition 
temperatures and the much larger $T_{MF}$.  Evidence 
for these fluctuations has previously been obtained in X-ray 
diffraction \cite{pretransflucts}, ARPES 
\cite{schaefer,schaefer2}, and tunneling experiments \cite{dai}.  In 
light of these strong fluctuation effects, the mean-field or BCS-like 
form of the temperature dependence of the CDW gap observed in X-ray 
diffraction \cite{fleming} and point-contact spectroscopy experiments \cite{monceau} 
remains puzzling. 

Finally, we comment on the temperature dependence of the scattering
rates ($\Gamma$) for the Drude terms, shown in the inset of Fig. 1. 
The $\Gamma$ value for the 
first (narrow) Drude term is partially determined by the HR
extrapolation of $R(\omega)$. However, its temperature dependence 
is strikingly similar to that of $\Gamma$ for the second (broad) 
Drude term, which covers an extended
spectral range going well beyond the range of the HR extrapolation, 
suggesting that the extrapolation is not responsible for the observed 
behavior. 
 As directly
indicated by the narrowing of
$\sigma_{1}(\omega)$ with decreasing temperature, the
scattering rates show a pronounced drop below
50 $K$ for both Drude terms and both polarization directions. As in 
the
two-dimensional $2H-XSe_{2}$ dichalcogenide systems \cite{dordevic}
this clearly indicates that some scattering channels freeze out when 
the long-range-ordered CDW condensate develops. 
On the other hand, 
short range ordered CDW segments present for temperatures $T>T_{2}$ 
can lead 
to additional
scattering of the ungapped charge carriers. The weak temperature
dependence of $\Gamma$ above $T_{2}$ reinforces the notion\cite{ong} 
that the (CDW)
resistivity anomalies are due to changes in the carrier
concentration (inset of Fig. 3a) and not in the lifetime.

In conclusion, we have performed the first complete and bulk sensitive optical
investigation of $NbSe_{3}$, the most important CDW system to show 
collective transport.   We identify the energy scales associated with 
the CDW gaps,
determine the fractional gapping of the Fermi surface, observe CDW 
fluctuation effects above the experimental Peierls transitions, and 
establish the suppression of the free charge carriers scattering in the broken
symmetry ground state.  

\acknowledgments
The authors wish to thank J. M\"uller for technical help, and R. 
Claessen, J. Sch\"afer and M. Grioni for fruitful discussions. This 
work
has been supported by the Swiss National Foundation for the
Scientific Research and by the NSF (DMR-0101574).

\newpage


\begin{thebibliography}{10}
\bibitem{comment}
Different fit
procedures with variable number of fit components were considered but yielded equivalent results. The
temperature dependence of the extracted parameters and above all the redistribution of spectral
weight unambiguously converge to a unique trend.
\bibitem{gap}
Because of the remaining Fermi surface and Drude weight the CDW single 
particle excitations should be correctly named as pseudogaps. For 
simplicity we call them CDW gaps.
\end{thebibliography}
\end{document}